\documentclass[a4paper, amsfonts, amssymb, amsmath, reprint, showpacs, showkeys, nofootinbib, twoside, floatfix]{revtex4-2}
\usepackage[english]{babel}
\usepackage[utf8]{inputenc}
\usepackage{upgreek}
\usepackage[colorinlistoftodos, color=green!40, prependcaption]{todonotes}
\usepackage{amsthm}
\usepackage{amsmath}
\usepackage{mathtools}
\usepackage{physics}
\usepackage{xcolor}
\usepackage{graphicx}
\usepackage[autostyle]{csquotes}
\usepackage[left=16mm,right=16mm,top=35mm,bottom=20mm,columnsep=14pt]{geometry} 
\usepackage{adjustbox}
\usepackage{placeins}
\usepackage[T1]{fontenc}
\usepackage{lipsum, babel}
\usepackage{url}
\usepackage{natbib,hyperref}
\hypersetup{colorlinks=true, urlcolor=blue, linkcolor=blue, citecolor=blue}
\usepackage{multirow, booktabs}
\usepackage{epstopdf}

\begin{document}

\renewcommand{\figurename}{FIG.{}}
\title{Low-energy electron-induced ion-pair dissociation to "Trilobite-resembling" long-range heavy Rydberg system }

\author{Narayan Kundu$^1$}
\author{Vikrant Kumar$^1$}
\author{Dhananjay Nandi$^{1,2}$}%
\email{dhananjay@iiserkol.ac.in}
\affiliation{$^1$Indian Institute of Science Education and Research Kolkata, Mohanpur 741246, India}
\affiliation{$^2$Center for Atomic, Molecular and Optical Sciences $\&$ Technologies, Joint initiative of IIT Tirupati $\&$ IISER Tirupati, Yerpedu, 517619,
Andhra Pradesh, India}





\begin{abstract}
We have studied electron-induced ion-pair dissociation dynamics of CO using the state-of-art velocity map imaging technique in combination with a time-of-flight-based two-field mass spectrometer. Extracting the characteristics for O$^-$/CO nascent atomic anionic fragments from the low energy (25 - 45 eV) electron-molecule scattering, first-time, we have directly detected the existence of S-wave resonated Trilobite resembling a novel molecular binding energy mechanism, as predicted by Greene \textit{et al.} \cite{greene2000creation}. The energy balance demands ion-pair dissociation (IPD) lie within a long-range (<1000 Bohr radius) heavy Rydberg system. Modified Van Brunt expression capturing the deflection of dipole-Born approximation is used to model the angular distributions (AD) for the anionic atomic fragments. The AD fits reveal that the final states are dominantly associated with $\Sigma$ symmetries and a minor contribution from $\Pi$ symmetric states that maps the three-dimensional unnatural oscillation of Born-Oppenheimer's potential.


\end{abstract}

\maketitle


\section{Introduction}
A Rydberg molecule is an electronically excited molecular system significantly different from the normal electronically excited atomic states. In Rydberg, an interaction exists between the positive ionic core and a partially removed excited electron in the quasi-classical Bohr atomic framework that attributes quantise energy levels by following the Rydberg quantum defect formula associated with the distributed ionic core \cite{kirrander2018heavy}. Each Rydberg series finally converges to a threshold ionisation energy which results in a more negligible energy difference between near threshold Rydberg states (have long lifetimes), although the principle quantum number shifted to its higher value. Thus, Rydberg molecules are those in which an atom has a very high angular momentum electron located far from its nucleus, and the resultant dynamics strongly depend on the system's reduced mass.

Rydberg states can be broadly classified into two types based on electron-perturber (neutral ground state atom) interaction strength. The first one is the direct molecular analogues of Rydberg atoms, conventionally known as ion-pairs or heavy Rydberg system (HRS) \cite{hrs_review,kirrander2018heavy,hrs_coherent,kirrandar_pra}. However, a negative ion is formed when the high momentum scattered electrons in HRS are charged-transferred to the ground state atoms and possess Coulomb dominant electron-atom polarisation potential as modelled by Kirrander \textit{et al.} \cite{kirrander2018heavy}. Cruciality is that HRS are responsible for generating high amplitude vibrational Rydberg states when an atomic anion replaces the Rydberg electron \cite{dressed_ipd_prl,kirrander2018heavy}. Thus, vibrationally excited HRS (A$^{+}$B$^{-}$) are approximated as a long-range collision complex between two opposite-charge fragments in the ionisation continuum, attributing ion-pair or di-polar dissociation spectroscopy.

Without cooling, the experimental study of these vibrationally excited Rydberg states within the ground state of HRS is strenuous due to fragile Frank-Condon factors and shakily weak electronic transition dipole moments (TDM) \cite{Markson_2016}, although the multi-photon ionisation technique has recently provided relatively easy access for directly probing the ion-pair state \cite{parker2008direct}. Here, we report the results of an electron-induced molecular super-excitation to its ionisation continuum, wherein the generosity of an HRS has been precisely recognised using a time of flight (TOF) based state-of-the-art velocity map imaging (VMI) study with nascent anionic fragments. We splendidly observed the Trilobite resembling the lowest order Born-Oppenheimer molecular binding mechanism when plotting the intensity distribution pattern for the O$^-$/CO atomic anionic fragments, which strongly maps to the report of Ultra-long Rydberg molecular system \cite{greene2000creation}; the second class of Rydberg associated with higher order partial waves $(L\geq 2)$.

Accounting the higher-order partial waves $(L\geq 2)$ in the electron-perturber $(e-B)$ interaction, recent report \cite{dressed_ipd_prl} reveals an ULRM $(A^{*}B)$ can also be mapped onto a effective ion-pair $(A^{+}B^{Q})$, where the perturber $(B)$ is dressed by a fractional charge Q which independent of internuclear separation $(R)$. However, a class of long-range bound ULRMs perceives when the ground state atom (perturber) is embedded within the electronic wave-function of a Rydberg-like atom \cite{bellos_trilobite,dressed_ipd_prl,eiles2019trilobites} with bond lengths on the order of thousands of Bohr radii, and kilo-Debye regime large dipole moment in where triplet electron-neutral (perturber) scattering dominant over singlet \cite{topical_rev_green,ulrms_molphy}. Such phenomena mainly emerged in low-energy electron-molecule scattering with fascinating spin structures briefly discussed by Fey \textit{et al.} \cite{ben2009observation,ulrms_molphy}. However, fragmented anions (Rydberg electron + ground state atom) accumulate a significant phase shift when scattered off and sequentially demand an energy shift proportional to the S-wave scattering length \cite{fermi1934sopra,omont1977theory,swave_scatter_prl,meshkov_swave,dressed_ipd_prl}. For vanishing magnetic moment vector ($\mu$=0) underline with $\Sigma$ symmetry association, this type of S-wave binds ion-pair novel bonding mechanism is commonly known as `$Trilobite$'. Similarly, the `$Butterfly$', `$Dragonfly$', `$Firefly$' and `$Gadfly$' PECs correspond to the P, D, F and G waves dominated scattering, respectively \cite{dressed_ipd_prl}. The oscillatory fringes in the PECs vanish for increasing value of M, and the pre-eminent Coulombic nature of $L\geq 2$ PECs becomes apparent. Thus, the existence of Trilobite directly depends on the binding energy that originates from such scattering phenomenon.

Considering Fermi's delineation of pseudopotential \cite{fermi1934sopra} with negative scattering length, Greene \textit{et al.} \cite{greene2000creation} predicted such kind of Trilobite-like novel binding energy mechanism in the lowest Born-Oppenheimer oscillatory potential curve for the long-range molecular Rydberg states. Before our report, direct observation of a trilobite like Rydberg bonding mechanism was still in search, although Bendkowsky \textit{et al.} \cite{ben2009observation} indirectly map to the trilobite observation extracting the S-wave scattering length, polarizabilities and lifetimes of the Rydberg molecules experimenting with Rubidium. However, the Trilobite resembling ion-pair formation has a deflected formation mechanism from that of Greene's report \cite{greene2000creation}.

We have reported the electron-induced ion-pair dissociation (IPD) dynamics of O$^-$/CO bound atomic anions using the wedge slicing technique \cite{kundu2021effect}, and we observed a class of highly excited long-range heavy Rydberg states that strongly maps with the novel Trilobite molecular binding functional. Using TOF-based fragment's mass analysis combined with state-of-the-art VMI technique, time-gated parallel slicing has been extracted with other cases and briefly discussed in our previous reports \cite{chakraborty2016dipolar,nag2015complete,nag2015fragmentation}. A spectroscopic study by simply applying the conventionally used time-gated parallel slicing technique has recently revealed the drawback for exaggeration of lower momentum ions mainly due to the inclusion of entire Newton spheres of diameter $\le$ parallel slicing time window \cite{nag_nccn,moradmand_wedge,moradmand2013_wedge}. Here, we implement and use the canonical time-gated wedge slicing technique, overcoming ion exaggeration's drawback with low momenta ions \cite{kundu2021effect} to study the IPD of O$^-$/CO. 

\section{Results and Discussions}

\begin{figure*}
    \centering
    \includegraphics[scale=1.08]{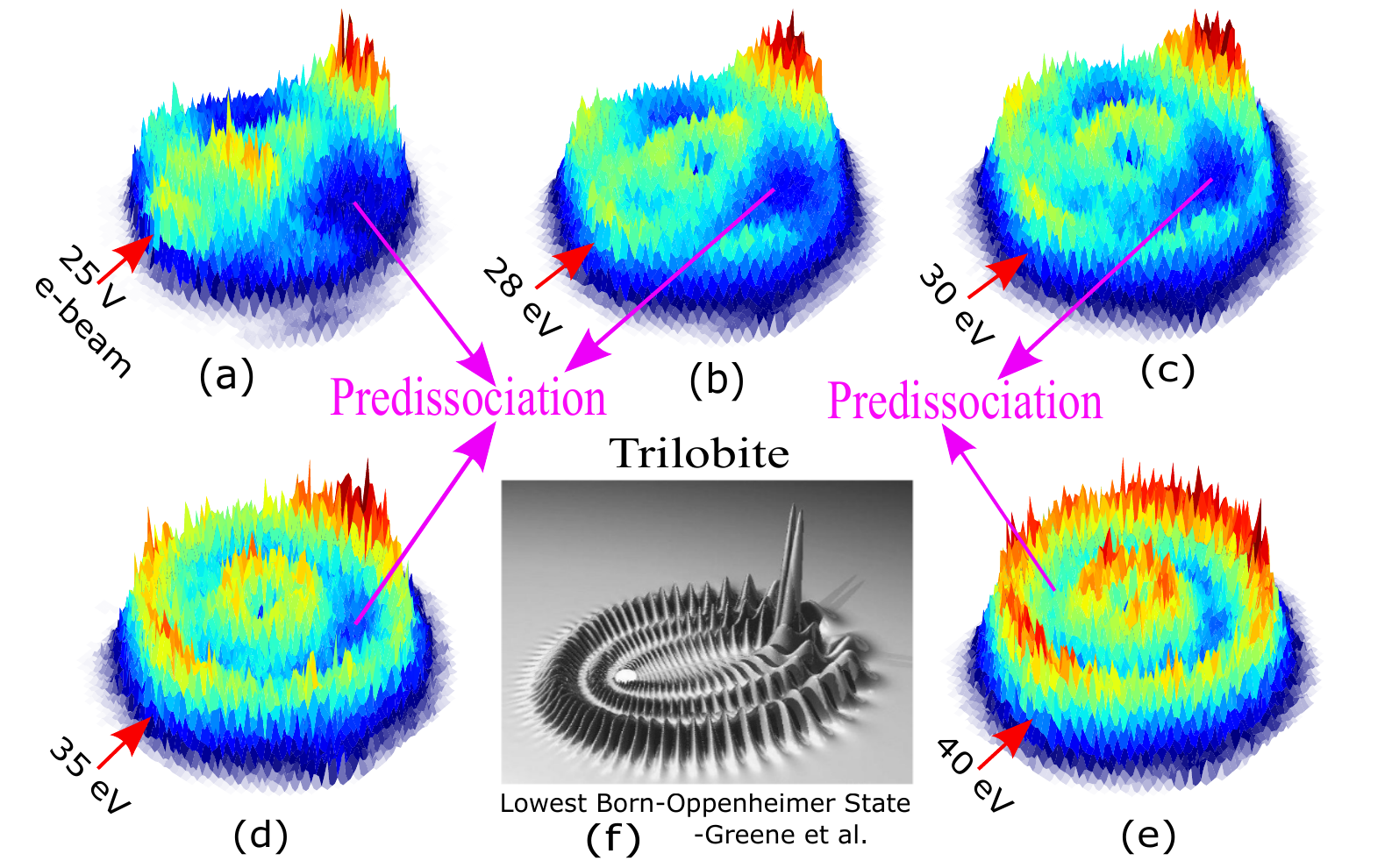}
    \caption{The surface plots (a,b,c,d, and e) depict the wedge slice images of O$^{-}$/CO extracted from electron-induced IPD dynamics for CO using state-of-the-art momentum imaging apparatus at incident electron energies 25, 28, 30, 35 \& 40 eV, respectively. (f) The lowest Born-Oppenheimer state as reported by Greene \textit{et al.} \cite{greene2000creation}. The red arrows indicate the induced electron beam direction labelled with the energy values.}
    \label{fig:vsi_co_ipd}
\end{figure*}

The FIG. \ref{fig:vsi_co_ipd} represents the wedge slice images of O$^{-}$/CO extracted from electron-induced IPD dynamics of CO using state-of-the-art velocity map imaging apparatus for incident electron energies at 25, 28, 30, \& 35 eV, respectively. To extract the exact emission function of O$^-$ products from the complex ion-pair bonding, we reroute the Abel inversion mathematical procedures by implementing the more practical wedge slicing method to capture the out-of-plane $\Pi$ symmetric transitional signatures (discussed later). All the velocity slice images (VSI); especially near 30 eV incident electron beam [Fig \ref{fig:vsi_co_ipd} (b,c)] strongly map with the Trilobite-resembling lowest Born-Oppenheimer oscillatory potential energy surface as predicted by Greene et al. \cite{greene2000creation}. A few rationales for our demand are grounded as follows.

The first question will arise, what is the correspondence between the wedge-shaped velocity slice images and Trilobite resembling Rydberg's electronic probability distribution which is presented using a surface plot in a cylindrical coordinate? An immediate answer exists within the state-of-the-art VMI technique and its proper visualisation through canonical time-gated wedge slice images containing anionic intensity distribution. In VMI, all the ionic fragments with a specific velocity vector are mapped onto a single point on the detector. As a result, fragments produced parallel to the detector plane (the central slice) are directly projected onto a two-dimensional position-sensitive detector without destroying the correlation for fragmenting. Here, the TOF is linearly mapped to the Z-axis for the cylindrically symmetric data structure (see the supplementary information), and the most probable central TOF maps the z=0 value, i.e. the x-y plane. Here, the anionic intensity anisotropy was detected on Cartesian's x-y plane for the cylindrically symmetric experimental data structure. Here, the wedge slicing is setting the camera of visualisation from the origin (x$_0$, y$_0$, z$_0$) of fragments emission, which enables one to extract the same fragments' emission function without exaggerating the lower momentum anions generating near the camera.

Secondly, why do the atomic anions distributions follow the same distribution fashion as Rydberg's electron probability distribution? Is this process limiting the Born-Oppenheimer approximation (BOA)? Here, the incident high energetic electrons beam is scattered by depositing its energy partially to the ground state CO molecules, lifting the molecule to its super-excited ionisation-continuum [(CO)$^{**}$], situated well above the molecule's first ionisation potential. Here, the excited Rydberg atoms within the dense background gases (atomic density >$10^{10}$ $cm^{-3}$ \cite{pfeiffer2008vacuum}) strongly allow electron-perturber (neutral Oxygen) triplet dominant S-waves scattering events. The scattered Rydberg electrons with relatively high momentum $\hbar k$ are weakly bound to the positive ion-core (C$^{+}$), which perturbs the more electro-negative Oxygen atom (via. polarisation), and finally charge-transferred to form the ion-pair with high vibrational amplitudes \cite{kirrandar_pra}. So, the simple electronic motion under the heavy mass nucleus is now replaced by complex ion-pair motion where O$^-$ anions revolving around the C$^+$ cations under the precedented attractive pseudo potential \cite{dressed_ipd_prl}, and preserves the BOA. Here, the Rydberg electron immediately follows the pathway of a highly excited neutral oxygen atom following the Condon approximation, and after dissociation, Rydberg's electron energy is reflected as the anionic energy. The effective motion can be framed out using Rydberg's mass-scaling factor ($\mu_{CO}$/$m_{e}$=12498), which marks a crucial role in controlling the energy of the Rydberg anion revolving around the positive ion-core \cite{kirrander2018heavy}. We push the anions toward the detector by breaking such attractive ion-pair pseudopotential of which bonding mechanism is reflected through the detected anionic intensity distributions.

Again, the wavelengths (935-468 a$_0$) corresponding to the induced electron beam energy (25-50 eV) conventionalised as the energy regime of the ion-pair imaging spectroscopy (IPIS) that strongly demands the ion-pair internuclear separation slickly lies in heavy Rydberg system (HRS; $\le$ 1000 a$_0$). We use a constant fraction discriminator (CFD) with a micro-channel plate (MCP) based delay line position-encoding detector, which detects nascent O$^-$ atomic fragments at about 4.8 $\pm$ 0.5 microseconds concerning with the start of electron beam pulse that agrees well with a recent report of Peper \textit{et al.} \cite{peper2020formation}; cation-anion plasma producing ultracold ion-pair would form even in an ULRMs (1000–2000 a$_0$) with microseconds order experimentally detectable lifetimes. Thus, O$^-$/CO has a minimum lifetime of 4.8 $\pm$ 0.5 microseconds; otherwise, it should not be detected.

\begin{figure}
    \centering
    \includegraphics[scale=0.42]{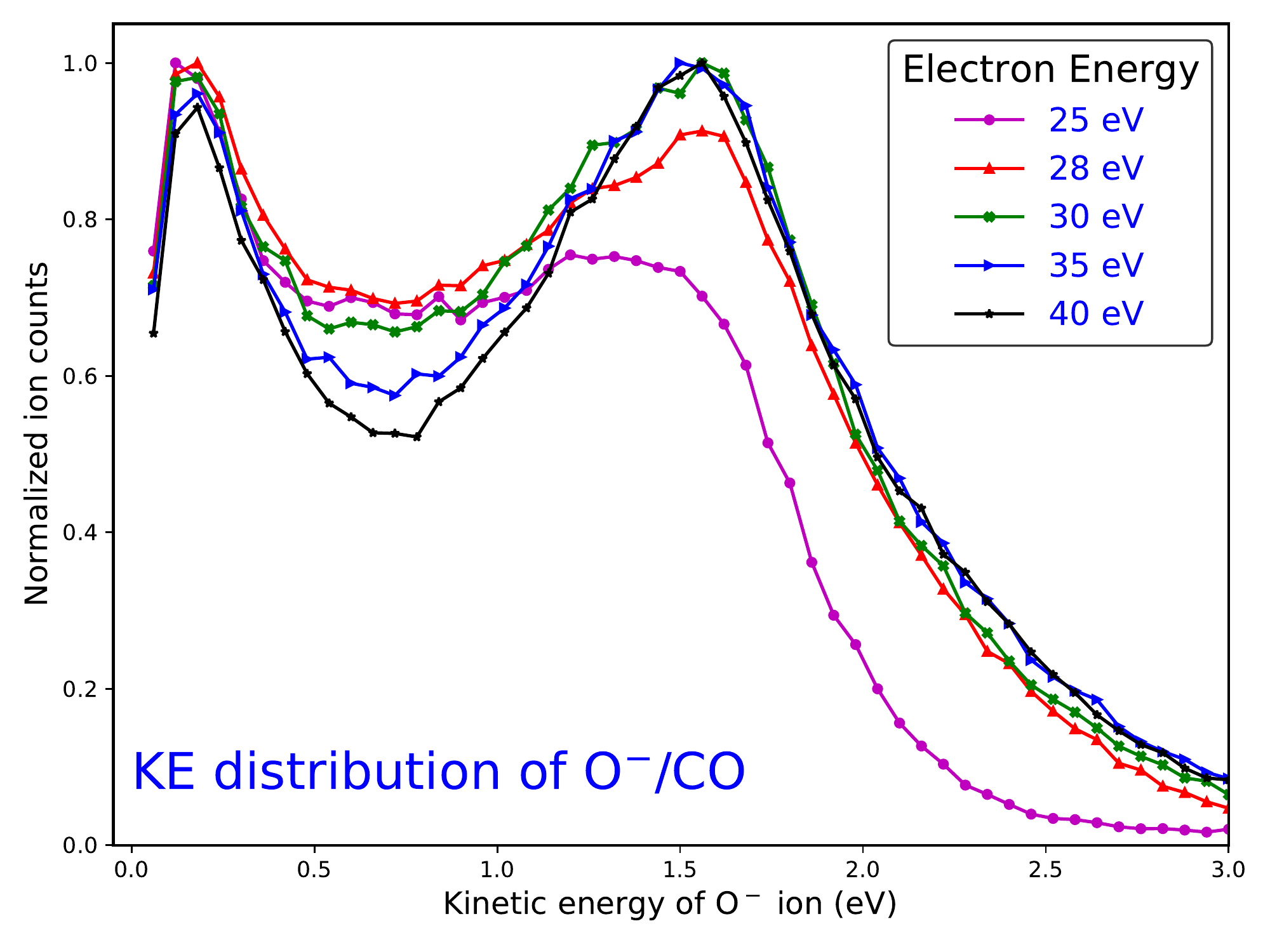}
    \caption{Kinetic energy (KE) distribution of O$^{-}$/CO nascent fragments at different incident electron energies as shown in legend with a different colour.}
    \label{fig:ke_co_all}
\end{figure}

However, we have determined the kinetic energy (KE) distribution of the O$^-$ by calibrating our measurement with the KE distribution for O$^-$/CO in the dissociative attachment process. When electron beam energy increases from 25 to 28 eV (near the threshold ion-pair production beam energy regime \cite{suits2006ion}), the KE distributions signify a shift for the most probable peak position of O$^-$ products that would correspond to direct access to the ion-pair state within the Frank-Condon transition limits. Beyond 28 eV (ion-pair imaging spectroscopic regime), directly accessible ion-pair states are out of the Frank-Condon transition regime that strongly demands de-excitation dynamics of a super-excited heavy Rydberg state. Whatever may make the excitations,  predissociation plays an essential role in reporting the dynamics for observing a continuum spectra behind the ion-pair characteristics spectrum. Generally, a continuum at a high wave number would correspond to usual dissociation, whereas the continuum, arising at energies well below the proper dissociation limit, is referred to as predissociation that can take place only when the Morse curves of a particular molecule in two different excited states intersect. One of the excited states is stable since it has a minimum in the curve (ion-pair state), and the other is continuous. However, a radiationless transition (faster than the molecular rotation: ~10$^{-10}$ s) occurs when the ion-pair state crosses over onto the dissociative continuum from its repulsive potential structure and attributes radiationless predissociation destroying all the rotational fine structures without affecting the vibrational structures \cite{banwell1994fundamentals}. However, destroying rovibrational structures, a complete continuum is reflected in the spectrum if the cross-over is faster than the vibrational time; the finest rationale for observing the predissociation continuum beneath the anion's high vibrational characteristics intensity spectrum (Trilobite Resembling) as shown using the wedge slice images in Fig. \ref{fig:vsi_co_ipd}.

We extracted the KE distribution from the wedge-shaped velocity slice images of O$^-$ that reveal the two most probable peak positions are positioned at 0.2 and 1.5 eV, as shown in FIG. \ref{fig:ke_co_all}. Thus, two distinct ion-pair states of CO exist in the Rydberg character-associated ionisation continuum. Without considering  Rydberg's mass-scaling, roughly, a quasi-classical Bohr formalism can be formulated for the O$^-$ formation from CO where Rydberg electron energy is charged and transferred to form O$^-$ anions. Then, 0.2 and 1.5 eV most probable KE of O$^{-}$ would correspond to the 8S and 3S (excluding the minor effect of other partial waves) molecular orbital of CO, respectively. The low-energetic threshold ion-pair production via the 8S Rydberg state of CO agrees with the report of Komatsu \textit{et al.} \cite{komatsu1995rotational}. However, the accurate energies of bound Rydberg quantum defect states ($E_{n}$) are connected with the principal quantum numbers(n) as follows \cite{kirrander2018heavy},

\begin{equation}
\label{eq:eq2_n}
    E_{n} = D_{C^+O^-}-\frac{[2hcR_{\infty}](\mu_{CO}/m_{e})}{2(n-\delta)^2}
\end{equation}

Where $[2hcR_{\infty}]$ signifies one atomic unit of energy, $R_{\infty}$ denotes infinite mass Rydberg constant for an electron, $\mu_{CO}$ and $m_{e}$ represents reduced mass of C$^{+}$O$^{-}$ complex and mass of the electron, respectively. The quantum defect, $\delta$, manifests the Rydberg level shifting from the pure Coulombic nature. Taking $E_{n}$=3.0 eV for maximum energy for O$^-$ bound state, and $D_{C^+O^-}$=19.8 eV for IPD threshold \cite{chakraborty2016dipolar} with negligible quantum defect $\delta$=0, we found that principle quantum number lies at value near 100, although this principle quantum number has a dependency with the background gas density \cite{topical_rev_green}. Here, the vibrational quantum number or the node counts ($\nu$) roughly replace the radial quantum number (n$_r$=$\nu$+1) associated with the electronic Rydberg states \cite{kirrander2018heavy}. Thus, the ion-pair attributes an effective pseudopotential, constructed to replace the complex effects of the motion for Rydberg's anions weakly attached to the cat-ionic core. Fermi's pseudopotential \cite{fermi1934sopra} also looks at the Rydberg electron as a quasi-free particle scattered with relatively high momentum $\hbar k$. The ion-pair potential, $V_{A^+B^-}(R)$, is expressed as follows.  
\begin{equation}
    V_{A^+B^-}(R) = D_{A^+B^-}-\frac{1}{R}-\frac{\alpha_{A^+}+\alpha_{B^-}}{R^4}
\end{equation}
The symbols used here are already discussed by Kirrander \textit{et al.} \cite{kirrander2018heavy}  where the attractive polarisation asymptote allows a finite number of bound states \cite{hrs_review}. Such full atomic potential eliminates core states manifesting pseudo-wave functions with significantly fewer Fourier nodes for the valence electrons that attribute plane-wave basis sets practical to use. 

\begin{figure}
    \centering
    \includegraphics[scale=0.42]{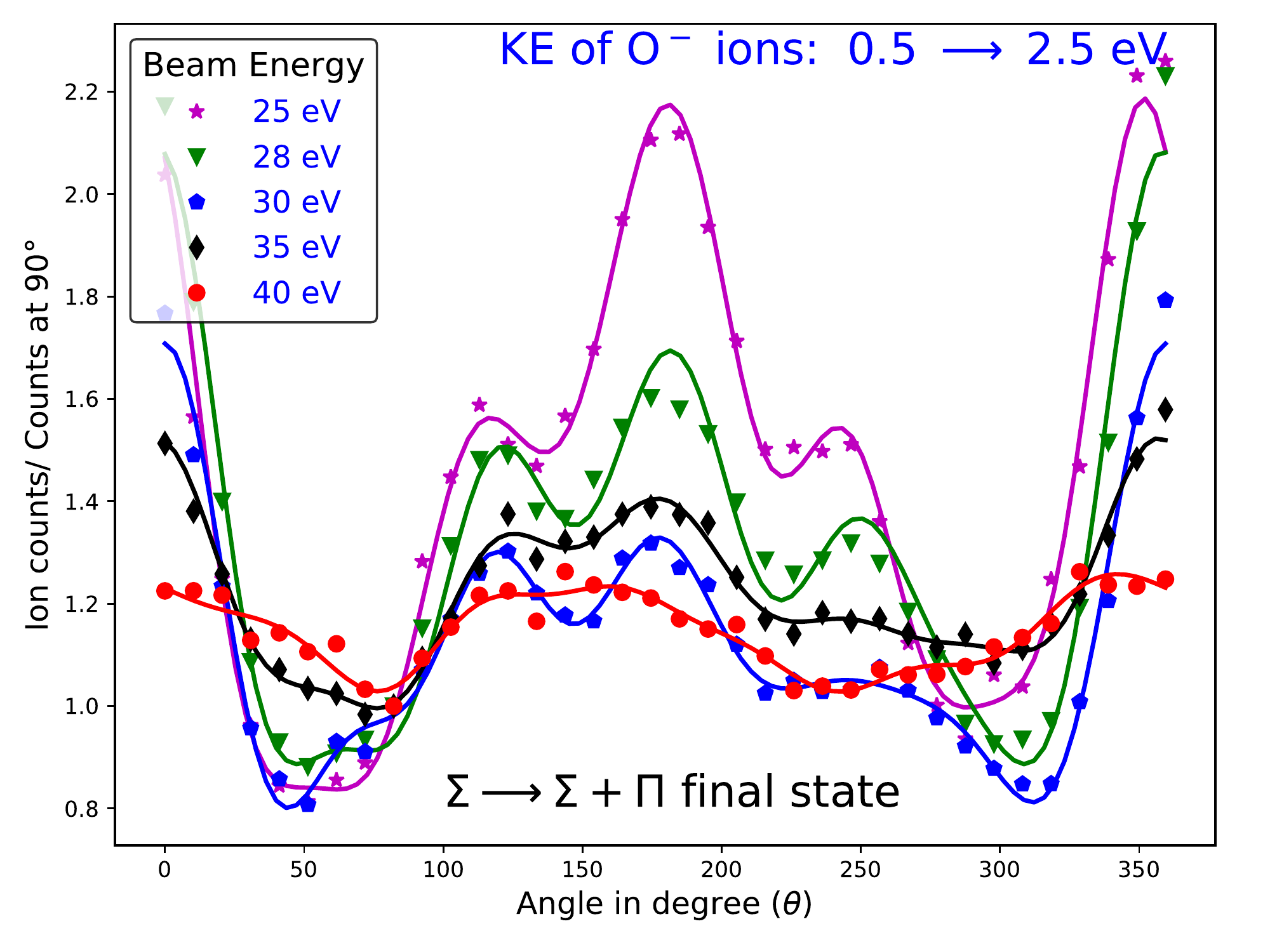}
    \caption{Plots represent the ADs of O$^{-}$ anions with KE from 0.5 to 2.5 eV as a function of electron beam energies and fitted using the mentioned modified Van Brunt expression. The final states are well associated with  $\Sigma + \Pi$ symmetric states, whereas $\Sigma$'s contribute more.}
    \label{fig:ad_ssp}
\end{figure}
\begin{figure}
    \centering
    \includegraphics[scale=0.52]{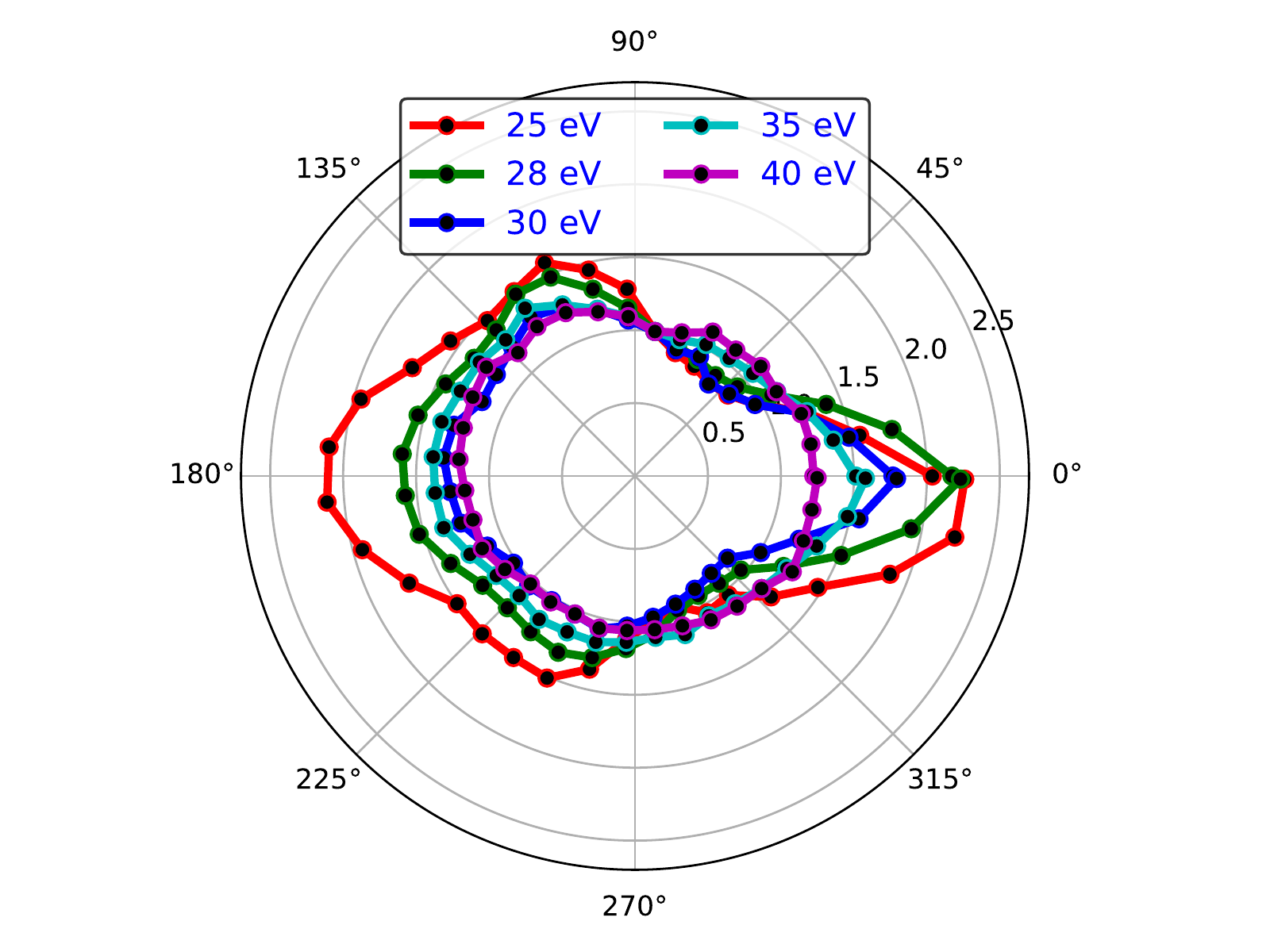}
    \caption{Polar plots for the ADs of O$^{-}$ anions with KE from 0.5 to 2.5 eV as a function of electron beam energies that strongly reveals the final states are dissociated via Vibrational frequency dependent S-wave resonated transitional dipole moments.}
    \label{fig:polar_plot}
\end{figure}

\begin{figure*}[ht!]
    \centering
    \includegraphics[scale=0.6]{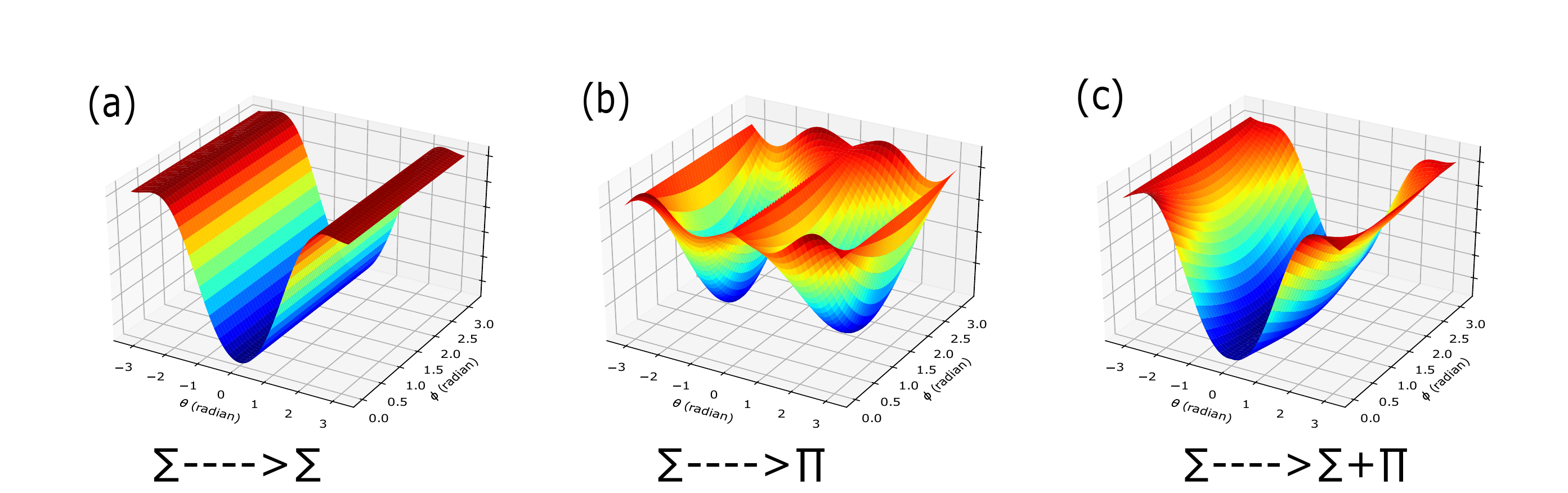}
    \caption{The angular intensity for the used fit function is plotted with the variation of azimuthal ($\theta$) and polar ($\phi$) angles. (a) The final states associated with $\Sigma$ symmetries signify oscillations in 2D (transitional moments) concerning to beam plane containing the variation of azimuthal ($\theta$) angle but zero transitional moments along with the variation of polar ($\phi$) angle. (b) Including $\Pi$ symmetries in the final states reveals transitional moments out of the beam plane along with the variation of polar ($\phi$) angle and (C) Existence of both symmetries, observed experimentally directly map the three-dimensional oscillation of the Born-Oppenheimer potential.}
    \label{fig:hetero_sy}
\end{figure*}

The KE integrated over 2$\Pi$ azimuths in the x-y plane, angular distributions (ADs) are also extracted from the wedge-shaped VSIs and presented in FIG. \ref{fig:ad_ssp}. To model the anisotropy in ADs, we have used Van Brunt \cite{van1974breakdown} formalism with the multiplication of an exponential of phase-shifts arising from the electron-perturber interactions ($e-B)$, as previously used by Nandi \textit{et al.}. According to the Van Brunt model, beyond the consideration of dipole approximation, the AD mainly depends on two functions; the first one is the Bessel function to calculate the primary effect of electron energy, and another one is a spherical harmonics [$Y_{l, \mu}(\theta, \phi)$] to account the perturber's wave function. We include another exponential  ($e^{i\delta_l}$) term to include the multipath scattering between the electron beam and loosely bound orbital electrons (triplet dominant), whose overlap with Spherical Harmonics generates oscillation within the 3D potential energy surface. This model directly maps the oscillation nature of the electron-perturber ($e-B)$ interaction \cite{dressed_ipd_prl}. During dissociation, the anionic wave functions in the vicinity of the cat-ionic core and should maintain the same symmetries for the fragment anions. The fit equation to determine the symmetries associated with ion-pair states is below.
\begin{equation}
    \rm{I}(\theta) = \sum_{\mid \mu \mid }\left\lvert \sum_{l = \mid \mu \mid}^{\infty} a_{l}i^{l} \sqrt{\frac{(2l+1)(l-\mu)!}{(l+\mu)!}}
    j_{l}(\kappa) Y_{l, \mu}(\theta, \phi)e^{i\delta_l}\right\lvert^{2}
\end{equation}
Here, j$_{l} $'s are the spherical Bessel function, $\kappa$ denotes the product of the momentum transfer vector, and the distance of closest approach between the incident electron beam and the centre of mass of the molecule, Y$_{l, \mu} $'s are the spherical harmonics, \textit{l} is the angular momentum of the electron and $\mu=\mid \Lambda_f - \Lambda_{i}\mid$, where $\Lambda_{i}$ and $\Lambda_{f}$ are the projection of angular momentum along the molecular axis for the initial and final states, respectively. Here, the summation over \textit{l} is responsible for the attachment of different partial waves.

During the ADs fitting using the mentioned modified Van Brunt expression \cite{chakraborty2016dipolar} as shown in Fig. \ref{fig:ad_ssp}, we notice that most of the transitional states are dominantly associated with $\Sigma \longrightarrow \Sigma$ symmetric selection ($\mu=0$); assigns anisotropy in beam plane only. However, an association of $\Pi$ symmetric transition ($\mu=1$) occurring perpendicular to the $\Sigma$s is designing three-dimensional Born-Oppenheimer oscillation governing the vibrational motion associated with the degenerate electronic manifold as depicted in Fig. \ref{fig:hetero_sy}. Thus, a $\Pi$ symmetric transition is associated with an out-of-plane (x-y) perpendicular transitional moment that demands the ion-pair potential is unnaturally oscillatory.

Interestingly, the anisotropy in ADs is decreasing with the beam energy increment, which can be explained using Zare's interpretation \cite{zare1967dissociation} and briefly discussed in our previous reports \cite{chakraborty2016dipolar}. The angular anisotropy decreases as presented using the polar plots of the O$^{-}$ fragments in Fig. \ref{fig:polar_plot} that depict S-wave resonated O$^{-}$ formation resembling the trilobite bonding mechanism in long-range molecular Rydberg states, of which Born-Oppenheimer potential curves are oscillatory in low-energy electron-atom scattering \cite{ben2009observation,fermi1934sopra}. The momentum transfer vector rotates more when beam energy increases, resulting in a loss of the anisotropy of the anion's high amplitude of vibration and demands interpretation using a fine rotational structure. We cannot measure the fine rotational structures, as the predissociation appearing here is destroying all the fine rotational structures.     

In generally, magneto-optical trap (MOT) based ultracold laser cooling technique \cite{njp_2020,peper2020formation,eiles2019trilobites} is used to probe the ULRM's high resolution spectroscopic characteristics extracting the rotational fine-structure characteristics where Rydberg electron as well as perturbed ground state atom's total angular momentum plays a crucial role in reporting the dynamics \cite{njp_2020,peper2020formation,shaffer2018ultracold,topical_rev_green,ulrms_molphy}. The molecule's electronic and vibronic scale properties are forced to freeze here with the help of high-frequencies detuned laser cooling, which might be a rationale for being incompetent to observe the trilobite bonding mechanism directly. Cruciality is that ultralong negative atomic ion in a single bound state does not attribute any cooling transition, \cite{njp_2020}, and hence MOT-based laser cooling is not necessarily needed to probe such singly bounded atomic anionic state properties at this electronic scale; a mandible rationale for directly observing the Trilobite resembling novel molecular binding energy mechanism simply using the low-energy inelastic electron scattering studying with highly polar diatomic CO molecule in the ultra-high vacuum condition. In a recent review, Greene et al. \cite{topical_rev_green} nicely pointed out that the Rydberg spectroscopy can be performed at room temperature in background gases with reasonable densities upon which the observed principle quantum number is strongly dependent.   

\section{Conclusions}
Thus, the experimentally observed Trilobite resembling ion-pair binding mechanism via Rydberg atoms excitation within the dense background gases (atomic density >$10^{10}$ $cm^{-3}$ \cite{pfeiffer2008vacuum}) is due to low-energy electron-perturber (neutral Oxygen) scattering events. We speculate that the scattering events dominate the triplet mode over the singlet. An experimental study with spin-polarised electron beams may confirm the demands beyond our scope of work. The KE and ADs analysis strongly supports the ion pair bonding functional between C$^+$ and O$^-$ maps Trilobite structure in the O$^-$ fragment's intensity distribution. The electron-induced ion-pair dissociation to CO preserves essential characteristics of the radiationless predissociation continuum spectrum.

However, Greene \textit{et al.} \cite{greene2000creation} proposed a multi-photon excitation mechanism to probe such long-range Rydberg states. Using the femtosecond pump-probed method combined with multi-photon absorbed ionization, Parker \textit{et al.} \cite{parker2008direct} probed the super-excited ion-pair dissociation dynamics of O$_2$ molecular states. Here, multi-photon absorbed directly accessible ion-pair is a quasi-resonant selective excitation to the repulsive limb of the stable ion-pair state. On the other hand, electron-scattered ion-pair is mainly formed due to the Rydberg atoms' excitation dynamics from a super-excited heavy Rydberg state in the ionisation continuum. Choen and Fano \cite{cohen_fano} pointed out nicely that the conventional selection rule, $\Delta l = \pm 1 $ for single-atom optical transitions doesn't hold for molecules with non-zero internuclear distance in the usual frequency range of spectroscopy as parity conservation will allow the Rydberg electron's angular momentum to transfer its energy to the rotational degrees of freedom. This formalism will drastically change for electrons with high angular momentum in their excited or ionised states. Rydberg electron always has a negligible overlap of its wave function with the ground state, can overcome the centrifugal barrier to perturb it, and can even approach the nuclei at higher and higher energies for higher and higher values of orbital quantum number \cite{cohen_fano}. Finally, we can say that electron-induced IPD dynamics to polyatomic molecules could provide a P-wave shape resonated Butterfly ion-pair bonding mechanism manifested with the gradient of Rydberg wave-functions probability density.

\section{Acknowledgements}

N. K. gratefully acknowledges the financial support from DST of India for the "INSPIRE Fellowship" program and expresses deep appreciation to Prof. C. Greene and his co-authors for reporting a few excellent articles on long-range Rydberg properties. We acknowledge Dr P. Nag and Dr D. Chakraborty for their contribution to electron-induced IPD dynamics. We gratefully acknowledge financial support from the Science and Engineering Research Board (SERB) for supporting this research under  Project No. "CRG/2019/000872".
\bibliography{bib_n_kundu}

\clearpage
\newpage
\onecolumngrid
\section*{\textbf{Supplementary}}
\section*{Data acquisition and analysis}
\begin{figure*}[h!]
    \centering
    \includegraphics[scale=1.1]{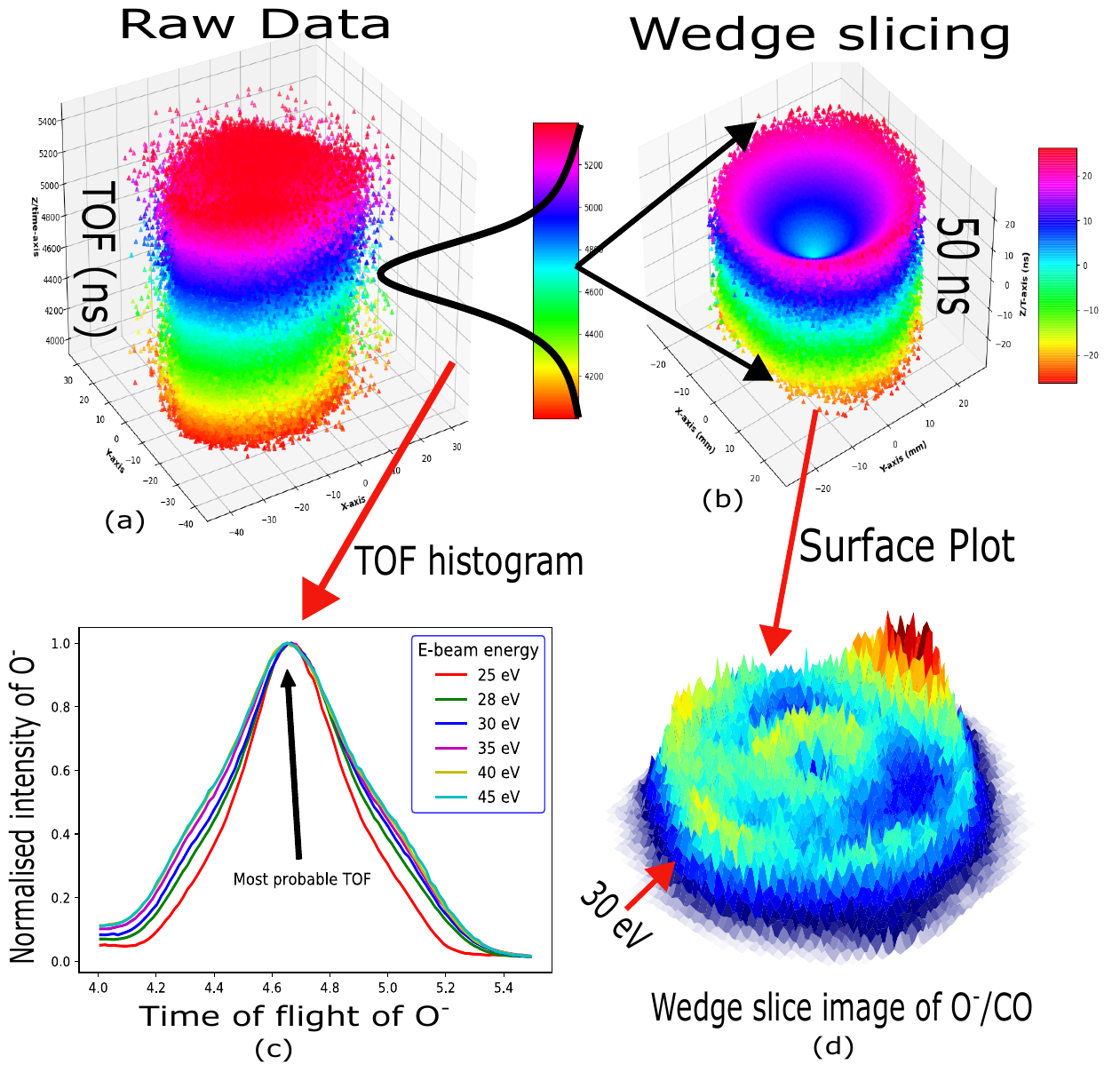}
    \caption{(a) MCPs and Hex-anode detected raw data, (b) Scatter plot of the wedge sliced data, (c) Time of flight histogram of the detected O$^-$ fragments, and (d) the surface plot of the wedge-shaped data.}
    \label{fig:raw}
\end{figure*}

We use a pulsed electron beam to collide inelastically with the effusive molecular beam having a diameter of 1mm of the pipeline. A 100 ns delay after the electron gun pulse is assigned for delayed extraction of the fragments, a rationale for observing better time slice resolution. The Pusher electrode-originated extraction pulse is then used towards detector assembly, maintaining the velocity map imaging conditions within the spectrometer.
Three z-stack micro-channel plates (MCPs) based on hex-anode detected raw data (x-position, y-position, TOF) provides cylindrical symmetry where the TOF (4200 ns to 5300 ns) maps with the cylinder's axis, as shown in FIG. \ref{fig:raw}(a). We found that the mean detected time scale of the O$^{-}$ nascent fragments is about 4.7 $\mu$s concerning pusher originated extraction pulse, and the full-width half maxima of the Gaussian-shaped detection is about 0.5 $\mu$s, as presented using the time-of-flight (TOF) histogram in FIG. \ref{fig:raw}(c). However, the practical lifetime of the O$^{-}$ anionic nascent fragments should be more than our detection time scale; otherwise, it should not be detected. Although, we are not accurately measuring the nascent fragment's lifetimes. Canonical time-gated wedge slice images are taken concerning the most probable central TOF value. Near this central TOF, the fragmented anions density is very high, providing better statistics. FIG. \ref{fig:raw}(b) represent the wedge slices of raw data, and most importantly, the wedge-shaped distribution of fragments is visible by the used colours. To extract the proper intensity distribution for O$^{-}$ anionic nascent fragments, wedge-shaped distribution is then plotted on a surface, as shown in FIG. \ref{fig:raw}(d).

\section*{Angular distribution fit parameter}
\begin{center}
\begin{table*}[!phbt]
 \centering
\caption{The model parameters reported here for the O$^-$/CO angular distribution as a function of electron energies with contributions from $\Sigma + \Pi$ final states. The a$_m$'s and b$_m$'s are the relative strengths of various partial waves for contributing to the $\Sigma$ and $\Pi$ final states, respectively. The $\delta_l$'s expressed in radian signify the phase differences concerning the lowest order partial wave responsible for the transition. $\beta_1$'s and $\beta_2$’sare the product of the momentum transfer vector and the distance of closest approach between the incident electron and the molecule's centre of mass for $\Sigma$ and $\Pi$ final states, respectively. Lastly, each m used in this caption is nothing but the whole numbers.}
\vspace{2mm}
\begin{tabular}{cccccc}

\hline
\hline
   \textbf{\vtop{\hbox{\strut Incident electron }\hbox{\strut \ \ \ \ energy (eV) }}\hspace{5mm}}    & \textbf{\vtop{\hbox{\strut Weighting ratio of the}\hbox{\strut different partial waves}\hbox{\strut $\hspace{10mm}a_0 : a_1 : a_2 : a_3$}\hbox{\strut $\hspace{10mm}b_1 : b_2  : b_3 : b_4$}}\hspace{4mm}}    & \textbf{\vtop{\hbox{\strut \ \ \ \ \  Phase differences in radian}\hbox{\strut \ }\hbox{\strut \ \ \ \ \ \ \ \ \ \ \ \ \  $\delta_{s-p}$, $\delta_{s-d}$, $\delta_{s-f}$}\hbox{\strut \ \ \ \ \ \ \ \ \ \ \ \ \  $\delta_{p-d}$, $\delta_{p-f}$, $\delta_{p-g}$}}\hspace{5mm}}  & \textbf{\vtop{\hbox{\strut$\beta$ parameter}\hbox{\strut }\hbox{\strut \ \ \ \ \ \ \ \ \ $\beta_1$}\hbox{\strut \ \ \ \ \ \ \ \ \ $\beta_2$}}\hspace{5mm}}   & \textbf{\vtop{\hbox{\strut$R^{2}$ value}\hbox{\strut }}\hspace{5mm}}\\
\hline

\multirow{2}{*}{25 }   &1.59 : 4.65 : 13.09 : 1.33    & 1.08, -0.13, 0.88  & 5.13 & \multirow{2}{*}{0.986}\\
                       & 4.75 : 1.61 : 4.01 : 2.91   & 1.44, 0.68, 3.14       & 2.76 & \\
\hline

\multirow{2}{*}{28 }   &2.44 : 4.28 : 6.02 :1.25   & 1.06, 1.02, 0.89       & 5.09 & \multirow{2}{*}{0.973}\\
                       &6.12 : 1.15 : 3.13 : 0.97    & 1.23, 0.90, 2.41       & 2.99 & \\
\hline

\multirow{2}{*}{30 }   &3.44: 2.47 : 3.06 : 2.96   & 0.29, 0.24, 1.68    & 1.83 & \multirow{2}{*}{0.980}\\
                       &2.31 : 0.99 : 0.89 : 0.90    & 0.44, 0.50, 1.08       & 3.45 & \\
\hline

\multirow{2}{*}{35 }   &2.22 : 3.33 : 0.24 : 1.09   & 1.87, 2.60, 2.15  & 2.71 & \multirow{2}{*}{0.979}\\
                       & 4.94 : 1.06 : 2.66 : 0.87       & 1.78, -0.89, 0.88       & 2.71 & \\
\hline

\multirow{2}{*}{40 }   &1.27 : 2.62 : 0.18 : 0.99    & 1.64, -0.08, 2.48  & 1.99 & \multirow{2}{*}{0.912}\\
                         & 4.17 : 0.71 : 2.61 : 1.81      & -1.73, -0.30, 2.52       & 2.45 & \\
\hline
\hline
\end{tabular}

\end{table*}
\end{center}
\end{document}